# Missing Dimensions: Integrating Human and Social Systems into Digital Twin Engineering

Francis Bordeleau
francis.bordeleau@etsmtl.ca
ÉTS | École de technologie supérieure
Montreal, Canada

Mark van den Brand
m.g.j.v.d.brand@tue.nl
Eindhoven University of Technology
Eindhoven, The Netherlands

## Abstract

Digital twins (DTs) have emerged as a key technology at the core of digital transformation, yet their engineering practice remains too narrowly focused on engineered and natural systems. This paper argues that four system dimensions must be explicitly recognized in DT engineering: *Engineered*, *Natural/Biological*, *Human*, and *Social*. Each dimension brings distinct properties, modeling requirements, and ethical obligations that fundamentally shape what a DT must represent and how it must be built. We further argue that as DTs extend into the Human and Social dimensions, the emphasis should shift from automated control toward decision support and occupant empowerment. We illustrate our arguments using a smart building as a running example and identify four open research challenges for multi-dimensional DT engineering.

*Paper submitted to EDTconf 2026 and currently under review.*

## 1 Introduction

Digital twins (DTs) have emerged over the past decade as a key technology at the core of digital transformation, now deployed across domains as diverse as industry [23], aerospace [13], agriculture [22], automotive [21], construction [4], earth monitoring [14], healthcare [16], smart buildings [25], smart cities [20], and telecom [2]. DTs differ significantly depending on their purpose, goals, and application domain [19]: a DT designed to improve energy efficiency in a smart building has little in common with one aimed at optimizing production in a manufacturing system or supporting personalized medical treatment. This diversity of purpose and context means that no single DT definition or engineering approach can be considered universally applicable.

Beyond differences in purpose and domain, the design and implementation of DTs vary significantly depending on the system dimensions they are associated with. We identify four main dimensions of systems that a DT may need to account for.

***Engineered*** – The *Engineered* dimension encompasses all types of systems, processes, and artifacts created by humans for various purposes.

***Natural/Biological*** – The *Natural/Biological* dimension covers systems that exist in the physical world independently of human intervention, driven by the fundamental laws of nature such as physics and chemistry, with biological systems forming a subset that consists of living organisms.

***Human*** – The *Human* dimension addresses the individual human occupants, users, or actors associated with a system, including their physiological states, behavioral patterns, cognitive characteristics, personal preferences, and individual needs arising from physical disabilities or neurodevelopmental conditions. Importantly, the Human dimension is concerned with the person as a cognitive and social agent, and is therefore distinct from the human body as a biological organism, which falls within the Natural/Biological dimension.

***Social*** – The *Social* dimension captures the networks of human agents, including individuals, groups, and institutions, that interact within a shared environment, driven by relationships, cultural norms, and communication rather than physical laws or rigid design specifications.

Together, these four system dimensions form a hierarchy of increasing abstraction and complexity that shapes that a DT must model, monitor, and support. Consider a smart building DT developed to improve energy efficiency and occupant comfort. Such a DT may initially focus exclusively on the engineered dimension by modeling the HVAC system, electrical network, and structural components. However, if occupant comfort is a primary goal, the DT must progressively incorporate the natural dimension to account for outdoor climate conditions and circadian light cycles, the human dimension to represent individual occupant profiles, preferences, and accessibility needs, and ultimately the social dimension to reflect how groups of occupants collectively use and shape the building environment. This progressive integration of system dimensions is not merely a technical challenge; it fundamentally changes what the DT is, what data it requires, what models it relies upon, and what ethical obligations its engineering must respect.

Based on our experience and a review of the DT literature, we observe that the concept of DT is too often reduced to its engineered or natural/biological dimension. **The human and social dimensions are frequently ignored or only superficially addressed, resulting in DTs that optimize physical systems while neglecting the people-centric purposes those systems are ultimately meant to serve** [6, 8, 9, 15, 26]. **We argue that this limitation does not reside primarily in the DT artifact itself, but in the process by which DTs are engineered.** Current DT engineering practices lack systematic guidance for identifying, integrating, and balancing the multiple system dimensions that a DT must account for, particularly as it extends beyond the engineered realm into human and social territory.

**This paper takes a position on the need to broaden both the definition and the engineering practice of DTs by explicitly recognizing and addressing all four dimensions.** While we use

a smart building as a running example throughout the paper, the framework applies equally to other domains including healthcare, smart cities, and industrial systems.

The main contributions of this position paper are threefold. First, we introduce a framework of four system dimensions that a DT may need to account for, Engineered, Natural/Biological, Human, and Social, and characterize the distinct properties, modeling requirements, and ethical obligations each dimension brings to DT engineering. Second, we argue that as DTs extend beyond the Engineered dimension, the emphasis should progressively shift from automated control toward decision support and human empowerment, and discuss the implications this has for DT engineering practice. Third, we identify open research challenges that must be addressed to support the systematic engineering of multi-dimensional DTs.

## 2 Automated control is not essential for DTs

To ground our discussion of system dimensions, we first clarify what we consider to be the essence of a DT. Many definitions of DTs have been proposed in the literature, typically reflecting the specific concerns, priorities, and constraints of the domain and context in which the DT is being developed and used [7, 10, 12]. Despite this diversity, a common conceptual core emerges: a DT is a dynamic digital replica of a physical object, process, or system, referred to as the *Actual Twin (AT)*[1], that is continuously updated with data from the AT and its environment[2] [11, 24], and that can interact with and influence the AT. A DT is defined by four foundational pillars: *purpose*, *services*, *models*, and *data*.

A DT must be grounded in a well-defined purpose with clear, measurable goals that guide its design, scope, and operation. The purpose determines what the DT needs to represent, at what level of fidelity, and for whose benefit. Aligned with this purpose, the DT delivers a set of services to its stakeholders, including monitoring, simulation, diagnostics, optimization, and what-if analysis, powered by a diverse set of modeling techniques spanning descriptive, predictive, and prescriptive categories [5]. These models may take many forms, including statistical models, differential equations, physics-based simulations, and AI/ML models. Underpinning the entire architecture is data: the DT is continuously updated with real-time or near-real-time data from the AT and its surrounding environment, ensuring ongoing fidelity and relevance.

The strategic goals of a DT follow from these pillars. First, a DT aims to provide actionable insight and foresight by delivering services that generate timely, evidence-based knowledge about the current state and anticipated future behavior of the AT, enabling informed decision-making by its stakeholders. Second, it supports safe and cost-effective analysis by enabling what-if scenarios, root-cause investigation, and other analytical methods to be conducted on the digital model rather than on the AT itself, avoiding the risks, costs, or disruptions that direct experimentation on the AT would entail. Third, a DT strives to improve the AT by using the knowledge and analysis it produces to influence the AT's behavior, whether through automated actuation, decision support to human operators, or recommendations to governing stakeholders, with the aim of optimizing specific aspects of its performance or outcomes.

**A key implication follows from the third goal: automated control is not an essential characteristic of a Digital Twin**. While it is often portrayed as the ultimate goal of DT, such as the categorization defined in [12], most established DT frameworks describe a capability spectrum ranging from basic monitoring and simulation, through diagnosis and prediction, to prescription and, at the highest level of maturity, automated control. The core and widely agreed-upon essential characteristics of a DT are the bidirectional data linkage between the AT and its digital representation, real-time or near-real-time synchronization of state, faithful representation of the AT, and the ability to reason about the AT through the digital model. None of these foundational characteristics inherently requires automated control. A DT that provides actionable insight through monitoring, simulation, and decision support to human operators is fully deserving of the designation, regardless of whether any automated actuation is present.

This distinction carries particular significance when extending DTs into the human and social dimensions. Automated control may be appropriate and desirable for the engineered dimension, for example by automatically adjusting HVAC setpoints in response to detected faults or energy demand signals. In the human dimension, however, occupant agency, personal preferences, privacy, and individual diversity, including physical disabilities and neurodevelopmental conditions such as ASD, must be respected. An automated system that unilaterally adjusts an individual's environment, however well-intentioned, may conflict with that person's sense of agency and comfort, or may fail to account for needs not yet captured in their occupant profile.

This leads to an important position for DTs that incorporate natural/biological, human, and social dimensions. **As DTs extend beyond the engineered realm, the emphasis should progressively shift from automated control toward decision support, recommendation, and occupant empowerment.** Rather than autonomously acting on the physical environment, a human- and socially-aware DT should surface insights, offer personalized recommendations, and present options, keeping human occupants meaningfully in the loop and in control of their own environment. This not only better respects human dignity and agency, but also acknowledges the inherent complexity and context-dependence of human and social systems, which are far less amenable to automated optimization than engineered ones.

## 3 System dimensions

As introduced in Section 1, we identify four fundamental system dimensions that a DT can be associated with, each characterized by its own properties, level of complexity, and degree of uncertainty. These dimensions are not mutually exclusive; in practice, a given AT will often span multiple dimensions simultaneously, and a DT must account for those dimensions that are relevant to its defined purpose and goals.

The **Engineered** dimension is the most tractable from a modeling perspective. Engineered systems are designed according to explicit specifications and governed by well-defined rules, making their behavior largely predictable and their state observable through sensors and instrumentation. In the context of a smart building, the HVAC system, electrical network, lighting infrastructure, and building management system all belong to this dimension.

[1] While DT definitions traditionally refer to a Physical Twin, we use the term Actual Twin to reflect the fact that DTs can be associated with different types of systems that are not exclusively physical, including processes, Cyber-Physical Systems (CPS), and socio-technical systems.

[2] The frequency at which the DT is updated depends on its purpose and goals, and may range from real-time synchronization to periodic batch updates.

**Table 1: Digital Twin Dimensions in a Smart Building Context**

| | Engineered | Natural/Biological | Human | Social |
|---|---|---|---|---|
| **Physical Entities Represented** | HVAC systems, elevators, electrical networks, structural components, lighting systems | Plants/green walls, urban heat effects, daylight cycles, air quality ($CO_2$, allergens), weather patterns | Individual occupants, their physiological states (stress, fatigue, comfort), personal profiles including disabilities and neurodivergent conditions | Occupant groups, teams, communities of practice, organizational units |
| **Sensors & Data Sources** | Energy meters, vibration sensors, BMS (Building Management System), actuator states | Weather stations, $CO_2$ sensors, air quality monitors, lux meters | Wearables (heart rate, skin temp), presence sensors, badge readers, calendar systems, occupant-declared preference profiles, companion apps | Meeting room booking patterns, collaboration tools, floor movement data, surveys |
| **Models & Representations** | BIM (Building Information Model), CFD (Computational Fluid Dynamics) models, thermal models, electrical load models | Bioclimatic models, circadian rhythm models, plant growth models | Comfort models (Fanger's PMV), cognitive load models, individualized occupant profiles (incorporating physical disabilities, sensory sensitivities, ASD/ADHD traits, mobility needs) | Social network models, space utilization models, organizational interaction graphs |
| **DT Services** | Predictive maintenance of equipment, energy optimization, fault detection | Adaptive daylight control, air quality regulation, seasonal energy adjustment | Personalized thermal/lighting/acoustic comfort, proactive accessibility management, sensory environment regulation, predictability support for neurodivergent occupants | Team collaboration support, space allocation, crowd flow management |
| **Feedback & Actuation** | Automated HVAC adjustments, elevator scheduling, fault alerts to facility managers | Blind/shading control based on sun position, ventilation boost when $CO_2$ rises | Personalized workspace recommendations, alert when environment exceeds comfort thresholds, automatic accessible routing, pre-emptive sensory adjustments (lighting, acoustics, temperature), advance disruption notifications | Reconfiguration of shared spaces based on team interaction patterns |
| **Challenges** | Model fidelity, sensor calibration, system interoperability | Stochastic variability of natural processes, seasonal dynamics | Privacy concerns, individual heterogeneity, subjective comfort perception, ethical handling of sensitive disability/neurodiversity data, ensuring occupant agency and consent | Group dynamics complexity, conflicting individual vs. collective needs |

The **Natural/Biological** dimension introduces a higher degree of uncertainty, as natural systems cannot be fully specified or controlled by design. Biological systems add a further layer of complexity, as their behavior is shaped not only by physical and chemical laws but also by the dynamics of life processes. In a smart building, this dimension encompasses outdoor weather conditions, solar radiation patterns, circadian light cycles, and the physiological responses of the human body to environmental stimuli.

The **Human** dimension presents a qualitatively different modeling challenge. Unlike systems in the previous two dimensions, individual humans are active agents whose responses to the same environmental conditions can vary widely and evolve over time. Capturing this variability requires not only physiological and behavioral data, but also careful attention to privacy, consent, and individual autonomy. In a smart building, this dimension becomes relevant when the DT seeks to personalize comfort settings, adapt accessibility features, or tailor services to individual occupant profiles.

The **Social** dimension is the most abstract and volatile of the four. Social behavior emerges from collective interaction among individuals, groups, and institutions, and is therefore particularly difficult to predict or model with precision. In a smart building, this dimension comes into play when considering how groups of occupants negotiate shared spaces, how organizational policies shape building use, how communities respond in emergency situations, or how collective norms influence energy consumption behavior.

Table 1 provides a detailed illustration of how these four dimensions manifest across various DT aspects in a smart building context, including the physical entities represented, the sensors and data sources involved, the models and DT services required, and the specific challenges each dimension introduces.

Beyond these four dimensions, it is important to explicitly acknowledge the role of the environment. Every AT exists within and interacts with an environment that may itself span multiple dimensions, including outdoor climatic conditions (Natural), surrounding infrastructure (Engineered), and the broader social and institutional context (Social). The environment is not merely a backdrop; it actively influences the behavior of the AT and must be accounted for when defining the scope and boundaries of a DT.

These dimensions form a progression of increasing abstraction and uncertainty, from machines and specifications, through the laws of nature and life processes, to the complexity of individual persons and emergent collective dynamics. The data and models a DT relies upon will differ substantially depending on which dimensions are in scope, and the associated uncertainty increases as the DT extends from the Engineered toward the Social dimension. Moreover, all dimensions of the AT and its environment are in flux and the DT should accommodate the drift in the corresponding data and models [1, 17].

## 4 Impact of system dimensions on DT engineering

DT engineering refers to the principled process of designing, developing, deploying, and evolving a DT throughout its lifecycle. Successful DT engineering requires careful consideration of three key properties: scope, fidelity, and evolution. With respect to scope, a DT is not a total replica of its AT but rather a focused abstraction that represents only the aspects of the AT that are relevant to its defined purpose and goals. Regarding fidelity, the required level of accuracy and precision is dictated by the DT's purpose, and not every use case demands the same degree of representational detail; over-engineering fidelity beyond what the goals require introduces unnecessary cost and complexity. Finally, with respect to evolution, a DT should follow an incremental development lifecycle, starting from a focused and manageable scope and growing in complexity over time. As the DT matures, its scope, fidelity, models, and services will naturally evolve in response to changing needs and a deepening understanding of the AT.

In practice, ATs are complex entities that can be decomposed into multiple aspects, and any given DT typically addresses only a specific subset of those aspects. A smart building, for instance, can be characterized along many aspects, including energy consumption, room occupancy, occupant comfort, waste management, and security and safety. The set of aspects covered by a DT will typically grow as part of the evolution process, driven by factors such as evolving user requirements, new data availability, and expanded operational objectives [5, 18].

Crucially, each AT aspect may be associated with one or more of the system dimensions introduced in Section 3. Consider a smart building DT focused on occupant thermal comfort. Such a DT may initially treat the building exclusively as an engineered system, controlling humidity, temperature, and ventilation through HVAC models calibrated against predefined benchmark values. As the DT evolves, however, it may need to incorporate more sophisticated models that account for the specific physiological characteristics and personal preferences of individual occupants, given that people are not uniformly comfortable at the same temperature. This extension moves the DT's scope progressively into the Human and, ultimately, the Social dimension. This progressive engagement with multiple system dimensions is a defining challenge of DT engineering, and one that current engineering practices are insufficiently equipped to address in a systematic manner.

Extending a DT across system dimensions also has significant implications for how DT engineering is organized as a collaborative activity. Each dimension typically requires different types of expertise: mechanical and systems engineers for the Engineered dimension, climate or life scientists for the Natural/Biological dimension, behavioral scientists and accessibility specialists for the Human dimension, and sociologists or organizational theorists for the Social dimension. The DT engineering process must therefore support effective collaboration among contributors with diverse backgrounds, enabling those responsible for different aspects and dimensions to work as independently as possible while contributing to a coherent and consistent DT [3].

A further challenge is that while a DT is ultimately a software artifact, most domain experts involved in its development and evolution do not have deep software engineering expertise. This creates a gap between those who understand the system being modeled and those who build and maintain the DT. Bridging this gap requires DT engineering approaches that provide appropriate abstractions, languages, and tooling, allowing domain experts to contribute meaningfully to the definition, validation, and evolution of the models and data relevant to their dimension, without requiring them to engage directly with the underlying software infrastructure. Model-driven engineering techniques are particularly well suited to address this challenge, as they support the use of domain-specific languages and visual notations aligned with the vocabulary and concerns of each expert community.

# 5 Research Challenges

We identify four main research challenges that must be addressed to enable the systematic integration of all system dimensions in DT engineering.

***Cross-dimensional collaboration and tooling***. A multi-dimensional DT requires contributions from experts across engineering, life sciences, behavioral science, and social science. Most domain experts in the natural/biological, human, and social dimensions are not software engineers, yet they must be able to define, validate, and evolve the models and data relevant to their area. DT engineering methods and tools must therefore provide appropriate abstractions and domain-specific languages that bridge the gap between domain expertise and software implementation.

***Multi-dimensional reference architecture.*** A reference architecture is needed to support the principled integration of aspects and services spanning multiple system dimensions. Such an architecture must accommodate heterogeneous data sources, diverse modeling paradigms, and independently developed services, while adhering to sound software engineering principles such as separation of concerns and locality of change. This is essential both to enable parallel contributions from multi-disciplinary teams and to support the incremental evolution of the DT as new dimensions and aspects are introduced.

***Heterogeneous data and model integration.*** Each system dimension relies on fundamentally different types of data and modeling techniques: physics-based simulations and sensor streams for the Engineered dimension, bioclimatic and physiological models for the Natural/Biological dimension, behavioral and preference models for the Human dimension, and social network and interaction models for the Social dimension. Integrating these heterogeneous sources and models into a coherent, interoperable DT while maintaining consistency and traceability across dimensions remains an open and significant challenge.

***DT driß detection and management.*** A DT must remain a trustworthy representation of its AT throughout its lifecycle. However, as the AT and its environment evolve, the DT can drift from reality through model drift, data drift, or changes in the scope of the relevant dimensions. This problem is particularly acute for DTs incorporating human and social dimensions, as these are inherently more volatile and harder to keep synchronized. Detecting drift must therefore be an integral part of DT operation, and its detection must trigger appropriate DT evolution to restore trustworthiness.

# 6 Summary

This paper argues that current DT engineering practices are insufficiently equipped to handle the full complexity of the systems that DTs are meant to represent. We introduced four system dimensions, Engineered, Natural/Biological, Human, and Social, each bringing distinct properties, modeling requirements, and ethical obligations that directly impact how a DT must be designed, developed, and evolved. We argued that automated control, while appropriate for the Engineered dimension, should give way to decision support and occupant empowerment as DTs extend into the Human and Social dimensions. Finally, we identified four open research challenges that must be addressed to make multi-dimensional DT engineering a systematic and principled discipline.

# Acknowledgments

Hossain Muctadir for the insightful discussions that have contributed to this paper.